\documentstyle[aps,floats,graphicx]{revtex}

\begin{document}

\newcommand{\bec}{\begin{center}}
\newcommand{\ec}{\end{center}}
\newcommand{\be}{\begin{equation}}
\newcommand{\ee}{\end{equation}}
\newcommand{\beqn}{\begin{eqnarray}}
\newcommand{\eeqn}{\end{eqnarray}}
\newcommand{\bet}{\begin{table}}
\newcommand{\ent}{\end{table}}
\newcommand{\bib}{\bibitem}

\wideabs{

\title{
%Mixing induced surface morphology development in Ti/Pt: a molecular dynamics study
Cooperative 
mixing induced surface roughening in bilayer metals: a possible novel surface damage mechanism
%Enhanced material transport in Ti/Pt due to low energy ion bombardment: a molecular   %dynamics study
}

\author{P. S\"ule$^1$, M. Menyh\'ard$^1$, K. Nordlund$^2$} 
\address{$^1$ Research Institute for Technical Physics and Material Science,\\
Konkoly Thege u. 29-33, Budapest, Hungary,sule@mfa.kfki.hu\\
$^2$ Accelerator Lab., Helsinki, Finland
}
%\email{sule@mfa.kfki.hu}

\date{\today}

\maketitle

\begin{abstract}
Molecular dynamics simulations have been used to study a collective atomic transport phenomenon
by repeated Ar$^+$ irradiations
in the Ti/Pt interfacial system. The ion-induced injection of surface atoms to the bulk, the ejection of bulk atoms to the top layers together with surface erosion is strongly enhanced
by interfacial mixing. This process leads to a dense interfacial material, and broadening of the interface region.
The process scales with the relative difference of the atomic masses. We find that 
surface roughening and interfacial mixing is strongly coupled via an 
enhanced counterflow material transport normal to the surface which might be a novel surface damage mechanism.
This cooperative phenomenon is active when the bilayer system is subjected to a high dose
ion irradiation (multiple ion irradiations) and leads to surface cavity growth.

{\em PACS numbers:} 61.80.Jh 61.82.Bg 68.35.-p\\
{\scriptsize 68.35.-p	Solid surfaces and solid-solid interfaces: Structure and energetics, Radiation effects on specific materials,
68.35.Ct 	Interface structure and roughness,
61.82.-d 	Radiation effects on specific materials,
61.82.Bg 	Metals and alloys,
61.80.Jh 	Ion radiation effects,
61.80.Az 	Theory and models of radiation effects 
}
\end{abstract}
}

\section{Introduction}

  Low-energy ion bombardment (around some keV) is a widely used technique in surface analysis, sputter cleaning, depth-profiling, micromachining, TEM specimen preparation etc. A tremendous experience on various effects of ion irradiation on solid surfaces has been amassed \cite{Gnaser}.  There are also several theoretical descriptions of the process especially if we consider homogeneous material. The most visible effect of the ion bombardment, which can be
 studied easily experimentally (e.g. by means of TEM, STM, AFM) is the change in the topography of the surface. It is generally accepted that the surface morphology produced by subjecting the surface to a given ion fluence is the result of competition between various processes resulting in smoothing and roughening. The roughening can produce somewhat periodic structures and/or rough structures which can be characterized by various scaling lows (a detailed reference of the experimental results are given in ref. \cite{Barabasi,Carter}).

  Molecular-dynamics (MD) simulations have also demonstrated the existence of the above mechanisms and proved to be capable to describe the development of various surface features during low energy ion sputtering and bombardment \cite{AverbackRubia,Mayr,Nordlund,Bringa,Koster,Beardmore}.

  Three categories of atomic mass transport can be classified when surfaces are subjected to ion beam irradiation \cite{Gnaser}:
(i) Atoms are ejected from the top layers, usually from the surface and srictly speaking this is what we call "sputtering" \cite{Barabasi,Carter}.
(ii) The displacement of atoms within their own layers creating vacancies, eroded surface, etc.
(iii) Intermixing between layers \cite{AverbackRubia}. 

  These transport phenomena usually lead to the change of the surface morphology
when the thermal spike \cite{TS} is located close to the free surface. It has been shown recently
that in this special situation the surface damage mechanisms 
could be chategorized to four separate mechanisms \cite{Mayr}: ballistic collisions, liquid flow \cite{AverbackRubia}, microexplosions \cite{Ghaly} and coherent displacements \cite{Nordlund_Nature}.
Recent extensive MD and experimental studies showed that the ion beam irradiation leads to surface
modifications that are more extensive than just simple sputtering \cite{Mayr}.
 Ion irradiation of interfacial metal bilayers are much less studied by MD \cite{Colla,Rasmussen,Sule} although the literature of interfaces in crystalline solids is rich \cite{Sutton}.

  In this paper we will report on MD simulation of ion irradiation induced alteration in the case of 4 Ti atomic layer/Pt substrate system. This system is specific in that sense that we place the interface close to the free surface ($\sim 10 \hbox{\AA}$) hence the thermal spike effects occur not only close to the free surface but also in the vicinity of the interface. The irradiation conditions were: ion energy 1 keV at grazing angle of incidence ($83^{\circ}$). Though the applied conditions can be considered gentle still strong mixing was observed. We will conclude that because of the presence of two unlike atoms enhanced material transport resulting in roughening and mixing occurs.
 We would like to show that 
  in the case of a sample made of thin layers of different atoms with large difference in the atomic mass, not ion sputtering causes the surface erosion but it is rather induced
by interfacial mixing.
Morover the mixing of atoms with different atomic mass and the cavitation at the surface is strongly coupled via the ion induced atomic transport normal to the surface leading to a {\em cooperative phenomenon}. This process might be a novel surface damage mechanism never reported before.

 We irradiate a Ti/Pt system using several ion impacts consecutively.
The reason for choosing such an irradiation mode is evident, since practically in all applications 
a series of ions hit the surface, even for low ion doses \cite{Menyhard1}.
We show through MD simulations of 1 keV Ar ions impacting the Ti surface of the Ti/Pt system
that near surface interfacial mixing  has a profound effect on the mass transport.

\section{Sample preparation and simulation technique}

Classical molecular dynamics simulations (MD) were used to simulate the ion-solid interaction
using the PARCAS code developed by Nordlund {\em et al.} \cite{Nordlund_ref}.
Here we only shortly summarize the most important aspects.
Its detailed description is given in  \cite{Nordlund_ref} and details specific to the current system is in a recent
communication \cite{Sule}.

  The sample consists of 37289 atoms for the interface (IF) system
with 4 Ti top layers and a bulk which is Pt.
The lattice constants for Pt is $a \approx 3.92 \hbox{\AA}$ and for Ti $a \approx 2.95$ and $c \approx 4.68$ \hbox{\AA}.
At the interface (111) of the fcc crystal is parallel to (0001) of the hcp crystal
and close packed directions are parallel.
The interfacial system as a heterophase bicrystal and a composite object of two different crystals with different
symmetry are created as follows:
the hcp Ti is put by hand on the (111) Pt bulk and various structures are probed
and are put together randomly. Finally that one is selected which has the smallest
misfit strain prior to the relaxation run.
The remaining misfit is properly minimized below $\sim 6 \%$ during the relaxation process so that the Ti and Pt layers keep their original crystal structure.
The corresponding Ti-Ti and Pt-Pt interatomic distances are $2.89$ and $2.73$ \hbox{\AA} at
the interface.
The Ti (hcp) and Pt (fcc) layers at the interface initially are separated by $2.8$ {\AA} and
allowed freely to relax during the simulations. The variation of the equilibrium Ti-Pt distance within a
reasonable $2.6-3.0$ \hbox{\AA} interval in the interatomic potential does not
affect the final results significantly.
We find
the average value of $d \approx 2.65$ \hbox{\AA} Ti-Pt distance in the various irradiation steps
and also in the nonirradiated system after a careful relaxation process.
We believe that the system is properly relaxed and equilibrated before and between the irradiation steps.
We found no vacancies at the interface or elsewhere
in the system after the relaxation procedure.
We visually checked and found no apparent screw dislocations, misorientations or any
kind of undesired distorsions (in general stress-generator dislocations) at the interface or elsewhere in the system which could result in
stressed induced rearrangements in the crystal during the simulations.
The temperature scaled softly down towards zero at the
outermost three atomic layers during the cascade events \cite{Nordlund_ref}.
The simulations are repeated in a much larger system consisting of $\sim 72000$ atoms in order to rule out any kind of size effects.
%------------------------------------------------------
The simulation cell is of particular interest because 
it has been shown \cite{Sule} that at the applied sputtering conditions (1 keV Ar$^+$ with grazing angle of incidence)
the thermal spike is typically formed at the interface and thus the emerging processes can be easily studied.

 The initial velocity direction of the
impacting atom was $7$ degrees with respect to the surface of the crystal.
Further details are given in Ref. \cite{Sule}.
  A series of up to $10$ ion impacts are employed in order to study the effects of multiple
ion irradiations, e.g. to simulate high dose effect.
Between impacts the simulation time was $\sim 20$ ps.
This time was found to be sufficient to reach a relaxed system after each ion impact (e.g. the temperature approaches zero). Thus direct interaction between the cascade processes of various irradiation steps are excluded.
 It should also be remarked that the damage rate is extremely high due to the high dose effect
and therefore some recovery process might occur between the irradiation steps.
Although we did not find significant recrystallization effects up to several hundred ps.
Thermally activated process might induce, however, further recovery in the system.
We have done relaxation simulations on the irradiated sample at 300 K and found no recrystallization of the amorphised
top layers in the irradiated zone up to several hundred ps. 
In any case, we do not think that the alloy phase would phase separate either at room temperature
or on a longer time scale. We will demonstrate in the Discussion section that the 
the alloy phase is energetically competitive with the phase separated system (based on the cohesive energies).

  To obtain a representative {\em statistics}, 
the impact position of the incoming ion is varied
randomly within a $10 \times 10$ \hbox{\AA}$^2$ area. 
We find considerable variation of mixing as a function of the impact position
of the recoil.
Using a $20 \times 20$ \hbox{\AA} irradiation area (with 80000 atoms or more) we checked that
a further increase of the irradiated area does not affect the final results heavily, except
that the larger the area, the larger the number of incoming ions needed to get the same extent of
interfacial mixing (IFM) and surface roughening. This is because individual ions with impact positions far from
each other do not result in the high dose effect ($\sim 10^{15}$ Ar$^{+}$ ions/cm$^2$) we are interested in in this article.
Impact positions with sufficient proximity to each other (that is within the characteristic area $\sim 5 \times 5$ \hbox{\AA}$^2$, $\sim 10^{15}$ ion/cm$^2$)
lead to the phenomenon called {\em mixing induced enhancement of surface roughening} (MIESR).
The overlap of these individually eroded surface areas might lead to the macroscopic phenomenon of 
sputter removal
when a sufficiently high dose of ions are bombarded to the surface in that case when
sputtering is the only active mechanism for surface damage.
However, the final morphology of the sputtered surface might also be affected by MIESR
if sufficiently large interfacial mixing occurs close to the free surface.

 We saw no further recovery process after 20 ps simulations although
we have extended several runs until 500 ps.
Thermally activated processes do not also play a significant role up to moderate
temperatures (room temperature or so). Also we set the quenching rate to a
value (5 K/Fs) which seems to be a reasonable choice.
Reasonable variation of the quenching rate does not affect
the cooling of the system and does not lead to the recovery of the original crystal order in the upper layers. 

%------------------------------------------------------
\begin{figure}[hbtp]
\begin{center}
\includegraphics*[height=4.5cm,width=6.cm,angle=0.0]{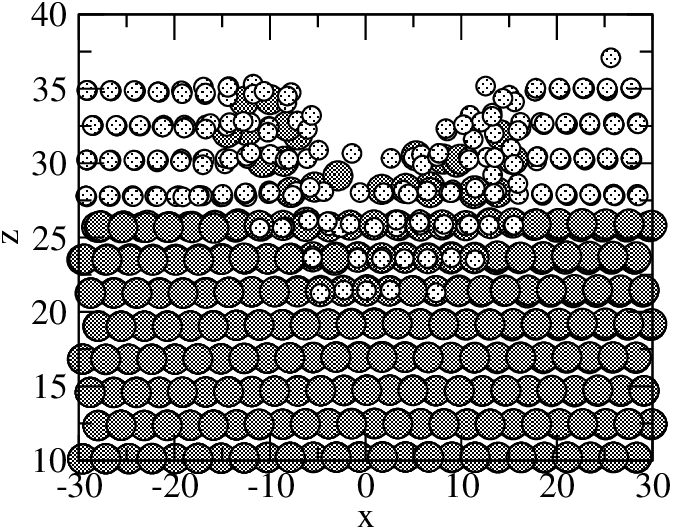}
\includegraphics*[height=4.5cm,width=6.cm,angle=0.0]{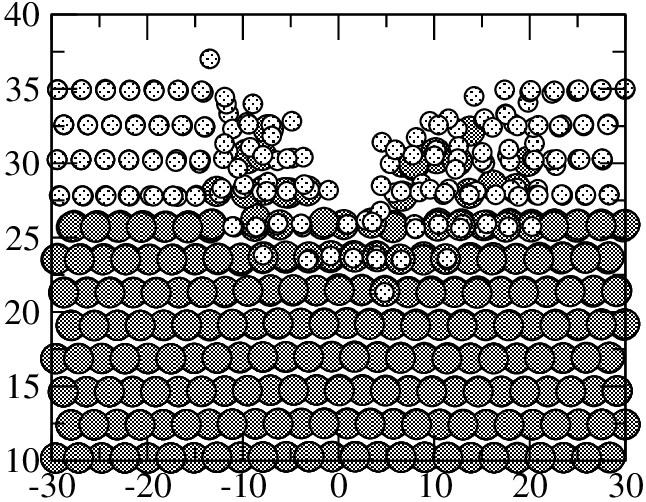}
\caption[]{
The cross-sectional view of Ti/Pt after 5 Ar$^+$ irradiations (upper figure) at natural
atomic mass ratio ($m_{Ti}/m_{Pt} \approx 0.25$).
The cross-sectional view of Ti/Pt after 3 Ar$^+$ irradiations at the mass ratio
of $0.06$ (lower figure).
The upper layer Ti atoms are shown with small circles.
The z-axis corresponds to the depth and the x-axis to the horizontal positions in \hbox{\AA}
.
}
\label{ptinti}
\end{center}
\end{figure}

%------------------------------------------------------

%------------------------------------------------------

\begin{figure}[hbtp]
\begin{center}
\includegraphics*[height=4.5cm,width=5.5cm,angle=0.0]{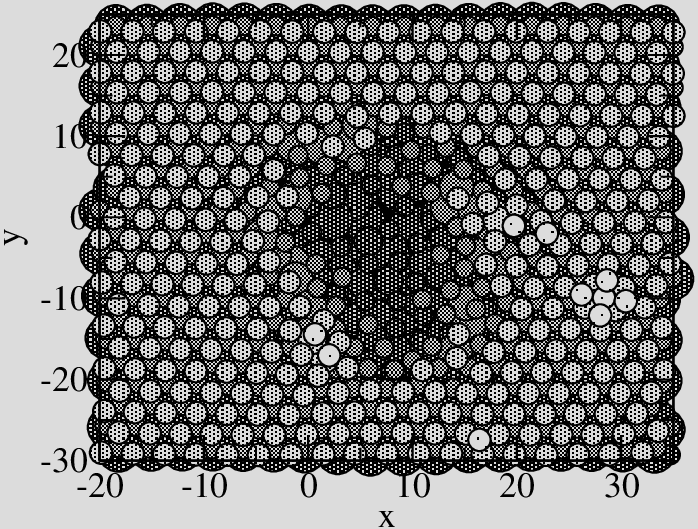}
\caption[]{
Cavitation (cratering) in the Ti/Pt system after the 5th bombardment.
The xy-plot of the Ti/Pt system seen from above (from the free surface).
}
\label{xy}
\end{center}
\end{figure}

%------------------------------------------------------

 During the simulations we observe extensive depression of the bombarded surface with respect to the nonsputtered surface. This depression did not result in a flat
 bottom, rather a rough surface had formed.
We estimate the extent of cavitation by counting the number of missing atoms in the cavity
using the same procedure introduced for vacancies \cite{Sule}.

 Vacancies were recognized in the simulations using a very simple analysis.
A lattice site with an empty cell was considered to be a vacancy, where the diameter of the cell sphere
is set to $\sim 2 \hbox{\AA}$ (the radius of a sphere around the relocated atom around its original position) which is about $60 \%$ of the average
atomic distance in this system.
We find this criterion is suitable for counting the number of vacancies.

 We used a many body potential, the type of an embedded-atom-method given by Cleri and
Rosato \cite{CR}, to describe interatomic interactions.
This type of a potential gives a very good description of lattice vacancies, including migration 
properties and a reasonable description of solid surfaces and melting \cite{CR}.
Since the present work is mostly associated with the elastic properties,  
melting behaviors, surface, interface and migration energies, we believe the model used should be suitable for this study.
 The Ti-Pt interatomic potential of the Cleri-Rosato \cite{CR} type is fitted to the experimental heat of
mixing of the corresponding alloy system. Further details are also given elsewhere \cite{Sule}.

 An atom is labeled mixed if it moved beyond the interface by more then $d/2 \approx 1.4 \hbox{\AA}$ along
the $z$-axis, where $d$ is the interfacial
Ti-Pt distance.
 The broadening at the interface (or interfacial mixing, IFM) is calculated as follows:
\be
b=N_{mix}^{-1} \sum_{i=1}^{N_{mix}} \Delta z_i,
\ee
where $N_{mix}$ is the number of mixed atoms and $\Delta z_i$ is the distance of the mixed atom $i$
from the interface ($z$-axis depth position).
$b$ is a useful quantity for comparison with measurements only in that case when $\Delta z_i$ is
larger then the interlayer distance. The reason is that the measured broadening is dominated
by those intermixed atoms which move beyond the interface by more than the inter-layer
distance (2nd layer mixed atoms). Using this condition we get a reasonable agreement with
Auger depth profiling \cite{Menyhard1}.

\section{Results and discussion}

 The detailed analysis of the mixing process in the Ti/Pt interfacial system
after a single ion irradiation process has been given recently \cite{Sule}.
 Here we restrict ourselfs to the study of the effect of multiple ion impacts
on mixing as well as on surface erosion.

%------------------------------------------------------
\begin{figure}[hbtp]                                                                        \begin{center}
\includegraphics*[height=4.5cm,width=6.cm,angle=0.0]{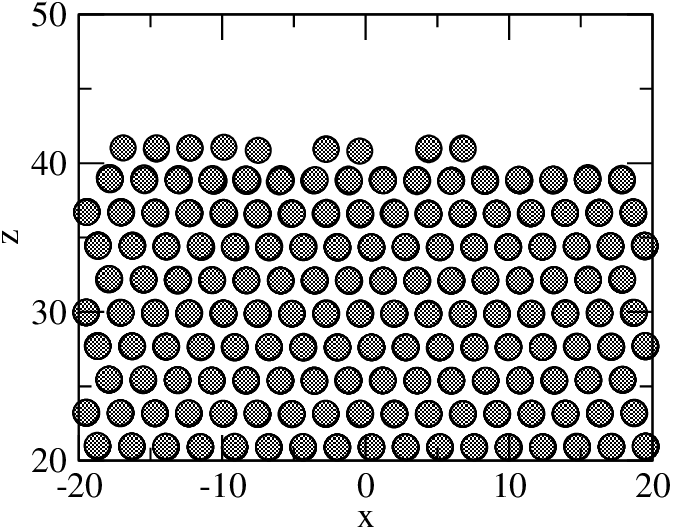} 
\includegraphics*[height=4.5cm,width=6.cm,angle=0.0]{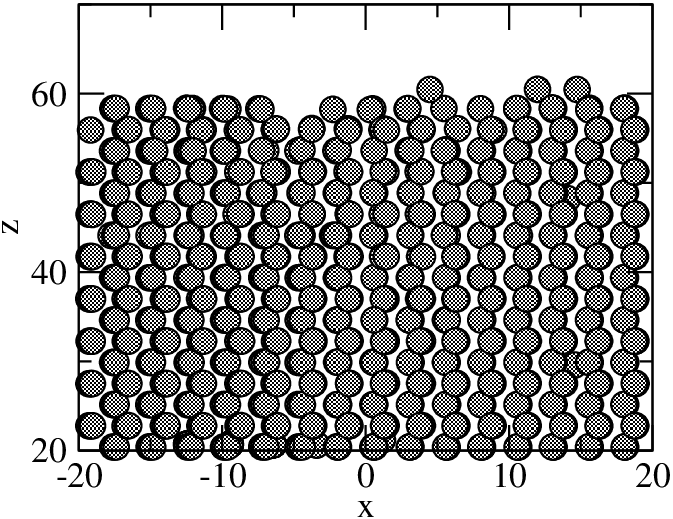}
\caption[]{                                                                                 The Cross-sectional view of pure hcp Ti (upper figure) and fcc Pt (lower) after 5 Ar$^+$ irrad                                                                                          iations.
}                                                                                           \label{ti}
\end{center}                                                                                \end{figure}
%------------------------------------------------------

\begin{figure}[hbtp]
%\begin{figure}[!t]
\begin{center}
\includegraphics*[height=4.cm,width=5.5cm,angle=0.0]{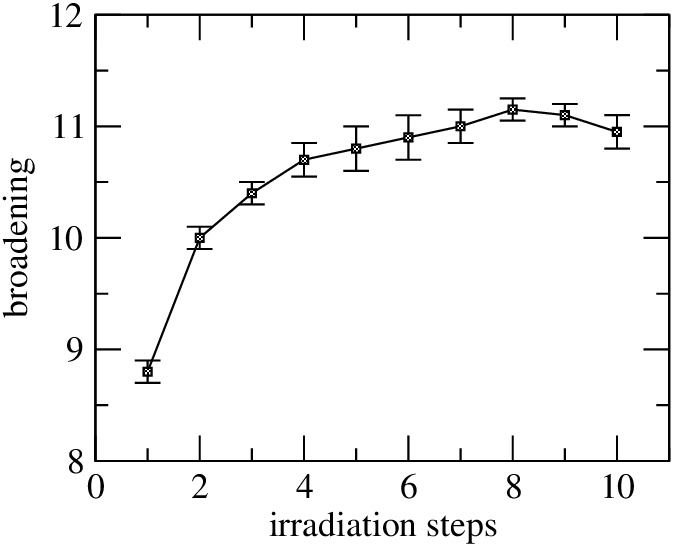}
\includegraphics*[height=4.cm,width=5.5cm,angle=0.0]{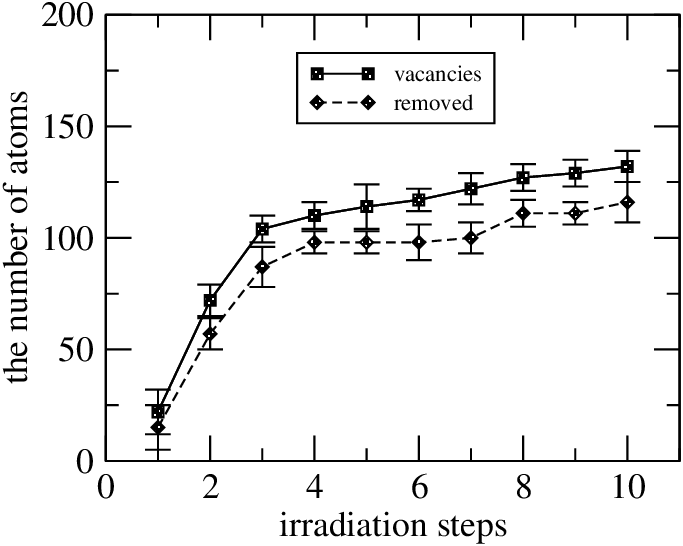}
\caption[]{
Upper figure: broadening at the interface ($\AA$) as a function of the irradiation steps.
Lower figure:
the number of removed (missing) atoms in the surface cavity as a function of
the irradiation steps. Note that most of the removed atoms are injected to the bulk and only
a small portion of them ($\sim 10 \%$) are sputtered to the vacuum.
The number of vacancies in the whole sample are also given.
}
\label{broad}
\end{center}
\end{figure}

 In FIGs ~(\ref{ptinti})-~(\ref{xy}) the $xz$ cross-sectional and $xy$ top plots of the Ti/Pt system 
are shown after the $5$th irradiation. It is evident that a considerable amount of material has been removed from
the irradiated area. Since  the height of the non-irradiated surface did not change apart from some adatoms the material removal 
produces a feature like a crater. It should be added, however, that the feature formed has a rough surface. 
The volume of the crater is much larger than which is expected considering only sputtered atoms. 
The number of sputtered atoms is less by an order of magnitude then the number of intermixed Ti atoms
which are pushed into the Pt bulk. These atoms cause the considerable interface broadening visible in 
FIGs ~(\ref{ptinti}) and ~(\ref{broad}).   
Indeed, the majority of removed atoms are the intermixed Ti atoms (injected to the bulk).
In FIGs ~(\ref{ti}) we show the corresponding pure components after the $5$th 
bombardment and see no apparent surface morphology development. This also shows that the degree of the material removal by sputtering is much weaker than the 
amount of the missing atoms in the case of the multilayered material.
%------------------------------------------------------
%------------------------------------------------------
Since the difference between the pure material and the Ti/Pt system is that in the case of the latter a two component system forms due to ion sputtering, 
we should conclude that the presence of the second component seriously affects the mass transport and thus most likely leads to morphology development as well. 
Let us check more closely the correspondence between IFM and mass transport (which is measured by the volume of the apparent crater).
{\em Our primary intention is then to point out that the enhancement of surface cavitation in Ti/Pt is due to
interfacial mixing.}
MIESR is strongly connected to a two-way material transport (with counterflow motion) which
might be understood as a radiation-induced flow of Pt
atoms to the surface and the
injection of Ti to the bulk.

  We see the strong enhancement of IFM due to the repeated irradiation of the surface in
FIG ~(\ref{broad}) where the broadening at the IF is plotted against the number of irradiation
steps (upper figure). 
Anomalously large intermixing in metal bilayers have recently been seen as well by Buchanan {\em et al.}
\cite{Buchanan}.
Surprisingly, we see the saturation of mixing (broadening) at the 5th step.
The magnitude of mixing at saturation shows no dependence on heat of mixing as it was obtained
in the first irradiation step \cite{Sule}. Hence
MIESR is not governed by thermal spikes. 

In the lower part of FIG ~(\ref{broad}) the number of "missing" atoms in the top layer (vacancies) is plotted 
against the number of irradiation steps. The number of missing atoms is proportional 
roughly to the volume of the surface cavity which
are mostly injected to the bulk.
Interestingly these features coincide with each other: the number of intermixed Ti atoms (the "missing" atoms) saturates
also at the $5$th step.

%------------------------------------------------------
\begin{figure}[hbtp]
\begin{center}
\includegraphics*[height=4.5cm,width=6.cm,angle=0.0]{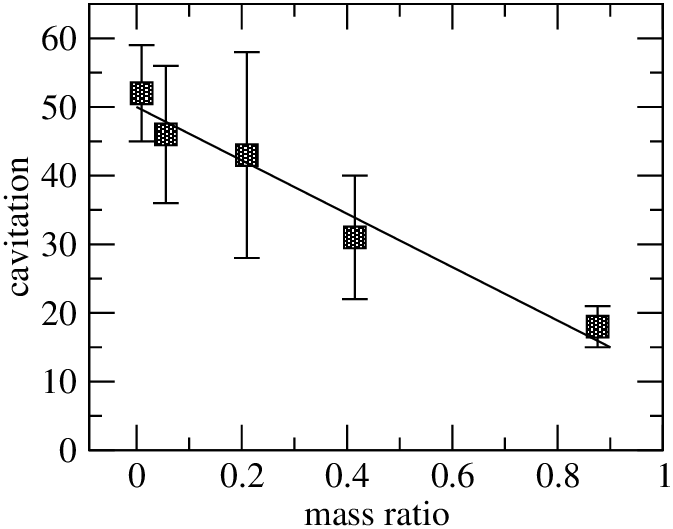}
\caption[]{
The number of removed atoms in the top layers (the extent of cavitation) as a function of the atomic mass ratio ($\delta=m_{top}/m_{Pt}$) in the third irradiation step.
The straight line is a linear fit to the data points, and the error bars denote standard deviations.
%Only a small portion of the removed Ti atoms ($\le 10 \%$) are sputtered (to the vacuum) and most of
%them are injected to the bulk.
}
\label{rem_massratio}
\end{center}
\end{figure}
%------------------------------------------------------

The volume of the cavity 
rapidly increases
in the first three irradiation steps and a crater similar to that of reported in earlier studies
by Bringa {\em et al.} at higher impact energies with single ion bombardment, forms \cite{Bringa}.
In these cases cratering is attributed to the high deposited energy close to the free surface at certain
nonchanneling directions.
In our case, however, after repeated irradiation of the surface, 
there seems to be a strong correlation between interfacial intermixing and material removal from the surface (material injection to the bulk).
Pt atoms intermixed to the top layers, however, do not occupy the empty sites in the cavity. Rather
they form a dense phase at the surface of the cavity.
The concentration of atoms in the irradiated region ($10 \times 10 \times 10$ \hbox{\AA}$^3$) increases from the initial $6.1 \times 10^{22}$ atom/cm$^3$
to $6.8 \times 10^{22}$ atom/cm$^3$ in the $5$th irradiation step.
The atomic concentration $n$ in TiPt alloy is $\sim 7.2 \times 10^{22}$ atom/cm$^3$, therefore the increase of
$n$ in the irradiated region indicates the formation of a broad alloy phase at the interface.
However, the saturation of the curves in FIGs ~(\ref{broad}) indicates that
the simultaneouss process of intermixing and surface cavitation is limited.
A possible control factor can be the increase in the number of the undercoordinated surface atoms which is always
unfavored. 
Another factor could be the saturation of the broadened interface.
No more intermixing of Ti atoms is observed possibly due to the denseness of the alloy phase
formed in the original Pt bulk which does not permit the further transport of Ti atoms to the
Pt bulk.
 Anyhow the maximal broadening of the IF seems to be limited to $b \approx 10 \hbox{\AA}$.
This restriction on broadening seems not to be affected by chemical forces
since it was found that the heat of mixing did not affect the IFM \cite{Sule}.

 In order to understand more fundamentally the mechanism of MIESR
we vary the atomic mass of the top layer constituent ($m_{top}$), keeping the same potential which can be
done easily in MD simulations \cite{Nordlund_ref}.
According to our expectations we find the suppression of IFM and surface cavitation when 
$m_{top}/m_{Pt} \rightarrow 1$.

%------------------------------------------------------
\begin{figure}[hbtp]
\begin{center}
\includegraphics*[height=4.5cm,width=6.cm,angle=0.0]{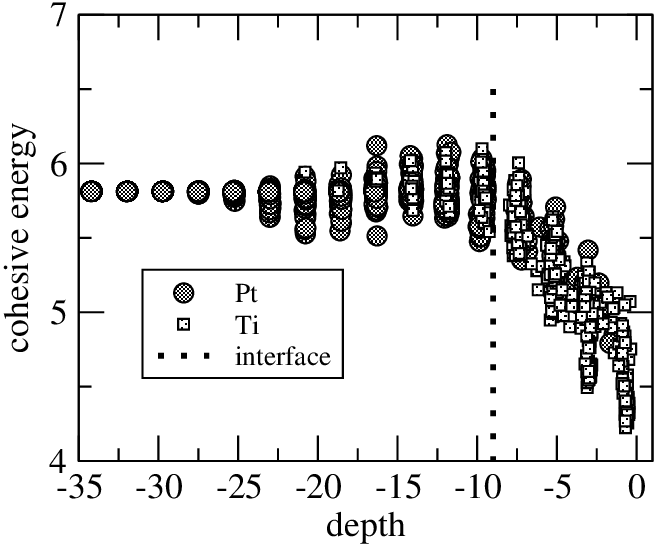}
\caption[]{
The cohesive energy (eV/atom) of Pt and Ti atoms as a function of their depth position ({\AA
}
) in
the
10th irradiation step.
The dotted line corresponds to
the depth positions of the interface and the free surface is at depth position $0$.
The cohesive energy in pure Pt and in Ti is $\sim 5.8$ eV/atom and $\sim 4.8$ eV/atom,
respectively, which should be compared with
the value of $6.3$ eV/atom in TiPt alloy.
}
\label{cohes}
\end{center}
\end{figure}
%------------------------------------------------------

  We plot in FIG ~(\ref{rem_massratio}) 
the number of Ti atoms which was removed from the top layers by MIESR (cavitation) 
as a function of the atomic mass ratio obtained from simulations 
in the third irradiation step.
In this step we obtained extensive mass transport especially for lighter atoms ($m_{top} \approx 2-11$ g/mol,
$\delta \approx 0.01 - 0.05$, see also FIG ~(\ref{ptinti})).
Notable feature is in lower FIG (1) the enhanced cavitation with small mass ratio in the third
irradiation step.  

 It can obviously be seen that the lighter the atoms of the top layers, the larger
 the extent of the mass transport. 
At the same time we see no obvious effect of heat of mixing on IFM \cite{Sule}.
It is therefore quite probable that IFM and surface roughening is governed primarily
by the atomic mass difference between the constituent atoms of various layers.
Mass transport from the surface layer is strongly enhanced for multilayers with light atoms in the
thin top layer.
This finding supports our previous conclusion \cite{Sule} that the ballistic mechanism governs the IFM.

This is interesting because we are not aware of similar results where the mass transport 
is strongly coupled to interfacial mixing.
Moreover the bulk process (mixing) and surface roughening are coupled via the enhanced
material transport normal to the surface leading to a cooperative phenomenon.
The importance of this finding is to point out that the surface of materials which include
more components close to the surface might behave in a different way than the pure components.
The surface erosion is governed not primarily by sputtering but rather by mixing.
A specific feature of this system is that we found no effect of chemical forces (heat of mixing) on
IFM \cite{Sule}. The process is governed by the large difference of atomic mass between Ti
and Pt.

  In order to understand the energy saldo of mixing we plot in FIG ~(\ref{cohes})
the cohesive energy (binding energy) of various atoms as a function of their depth position. The data points are collected from
the $10$th irradiation step.
The atomic cohesive energy $E_{coh}$ is averaged with that of every atom's nearest neighborsrs
to reduce the atom-level scatter in the values.
%to keep the atomic values at or below the bulk maximum in the TiPt alloy ($\sim 6.3$ eV).
It can obviously be seen that the cohesive energy of Pt atoms decreases when the atoms move towards the 
surface while Ti atoms gain energy when mixed to the Pt bulk.
%Therefore the driving force of mixing might also be the dramatic cohesive energy increase of the Ti atoms.
The energy gain per Ti atom is $\sim 1$ eV on average while approximately the same energy loss occurs for the
intermixed Pt atoms. 
%with previous studies \cite{Cheng}.
Above the depth position of $\sim -30$ {\AA} the widening of the cohesive energy "window" can also be seen for Pt. 
This is due to the formation of an alloy phase at the broadened interface.
Close to the free surface no features of alloy phase formation can be seen, e.g. the 
$E_{coh}$ values are rather close to the atomic value of pure Ti.
Also the $E_{coh}$ of the intermixed Pt atoms decreases at the surface which is due to their
decreased coordination number ($<12$) with respect to the bulk value of $12$.

 Finally it should also be mentioned that the ion induced surface cavity growth might have many common
features with the growth of cavities found on boundaries in stressed polycrystals or with cracking and in general with various fracture processes at heterophase interfaces \cite{Sutton}.
This is because we do not rule out the possibility of a cavity growth mechanism induced by a
stress field generated by the thermal spike. This is, however, beyond the scope of the
present article and needs further studies.

\section{Conclusions}

  We have shown that the mass transport for the studied double layer system is different from that of the corresponding pure elements. We have also shown that the observed apparent cavitation is in correlation with the interfacial mixing and it increases with increasing mass difference of the constituents. 
%The effect of the cohesive energy on cavitation is less obvious and needs further studies.
The most important consequence of this finding is 
that a reliable theory describing the surface morphology development of a system containing two components should also account for the mixing near the surface. Presently available theories are missing this point.
This novel surface damage mechanism might be applicable for those multicomponent systems (e.g. thin films) where the material is subjected to low energy ion-irradiation
and the interface is close to the free surface.
The mechanism is regressively reliable for systems with mass ratio approaching $1$.

\section{acknowledgment}
{\small
This work is supported by the OTKA grants F037710
and T30430   
from the Hungarian Academy of Sciences and from EU contract no. ICAI-CT-2000-70029.}

\vspace {-0.5cm}

%\section{Conclusion}

\end{document}